\documentclass[twocolumn]{revtex4-1}

\usepackage{graphicx}
\usepackage{dcolumn}
\usepackage{bm}
\usepackage{setspace}

\usepackage[utf8]{inputenc}
\usepackage[T1]{fontenc}
\usepackage{mathptmx}
\usepackage{etoolbox}
\usepackage{amsmath}

\makeatletter
\def\@email#1#2{%
 \endgroup
 \patchcmd{\titleblock@produce}
  {\frontmatter@RRAPformat}
  {\frontmatter@RRAPformat{\produce@RRAP{*#1\href{mailto:#2}{#2}}}\frontmatter@RRAPformat}
  {}{}
}%
\makeatother
\begin{document}

\preprint{}

\title{
Extrapolating Collisional Rate Coefficients for Molecules of Astrophysical Interest Using Neural Networks
}

\author{Yi\u git Ergen}
\affiliation{KU Leuven, Department of Chemistry, Celestijnenlaan 200F, 3001 Leuven, Belgium.}
\author{J\'er\^ome Loreau}
\email{jerome.loreau@kuleuven.be}
\affiliation{KU Leuven, Department of Chemistry, Celestijnenlaan 200F, 3001 Leuven, Belgium.}

\date{\today}

\begin{abstract}
State-to-state rate coefficients for inelastic molecular collisions are essential inputs for non-local thermodynamic equilibrium radiative transfer models, yet their computation using quantum scattering methods rapidly becomes prohibitive for large molecules and highly excited rotational states. We investigate the ability of artificial neural networks to extend existing datasets of collisional excitation rate coefficients without performing additional scattering calculations. A multilayer perceptron model was trained using molecular quantum numbers, level energies, energy gaps, and kinetic temperature as descriptors, and optimized on logarithmic rate coefficients. The model was tested on a diverse set of collision systems ranging from simple diatomic molecules (CO, SiO) to open-shell species (SO), small asymmetric tops (H$_2$O, SO$_2$), and complex organic molecules (benzonitrile and vinyl cyanide) colliding with H$_2$ molecules or He atoms. Extrapolations were performed toward higher rotational levels, higher temperatures, and combinations of both. For most systems, predictions of rate coefficients involving previously unseen rotational states yielded mean error factors between 1.1 and 1.5, with the dominant rate coefficients generally reproduced more accurately than weaker transitions. Temperature extrapolations proved particularly successful, while combined extrapolations in temperature and rotational excitation remained accurate to within a factor of two in most cases. The results reveal that prediction quality depends on the interplay between dataset size, temperature range, molecular complexity, and the presence of clear collisional propensity rules. Overall, the approach demonstrates that machine learning can provide a practical route to extending collisional datasets for astrophysical applications, especially for large molecules for which additional quantum-scattering calculations are computationally demanding.
\end{abstract}

\maketitle

\section{Introduction} {\label{introduction}}

Machine learning (ML) methods have recently found an increasing number of applications at the intersection of Physics and Chemistry. Examples include models of interatomic potentials and forces with applications in (classical) molecular dynamics \cite{Chmiela2018r,Deringer2021}, the construction of global or local potential energy surfaces using artificial neural networks (ANNs) \cite{Li2015,Kolb2016} or Gaussian process regression \cite{Qu2018,Sugisawa2020,Konings2023} for applications to spectroscopy \cite{Kamath2018} and quantum or quasi-classical dynamics \cite{Krems2019}, including non-adiabatic processes involving excited electronic states \cite{Chen2018,Westermayr2019}, while other approaches have focused on improving reaction prediction \cite{Meuwly2021_MLRev} or on the direct prediction of reaction rate coefficients \cite{Laude2018,Houston2019,Komp2020}.

The present work focuses on using ANNs to predict state-to-state rate coefficients in molecular collisions for high rotational levels and high temperatures based on datasets obtained with quantum dynamics methods. Inelastic collisions play a crucial role in the interpretation of molecular spectra from a wide variety of astrophysical environments such as the interstellar medium, since the local thermodynamic equilibrium (LTE) assumption is not necessarily fulfilled as a consequence of the low local gas densities. The accurate modelling of molecular spectra demands taking into account the balance between radiative and collisional processes through a radiative transfer model \cite{Roueff2013}, which in turn requires the rate coefficients for the collisional excitation of the relevant molecules by the background gas composed mainly of H$_2$ molecules and He atoms.

The computation of inelastic rate coefficients follows a well-established paradigm. In a first step, the potential energy surface (PES) of the colliding system is constructed by means of quantum chemistry methods. The dynamics on this PES is investigated to obtain state-to-state cross sections, which are used to obtain temperature-dependent rate coefficients by averaging over a Maxwell-Boltzmann distribution of collision energies. 
The most accurate approach to the dynamics is the fully quantum-mechanical close-coupling method \cite{Arthurs1960,Green1976b} in which the time-independent Schr\"odinger equation is solved in a basis of radial and angular functions. However, the cpu time and memory requirements for solving the coupled equations scale as $N^3$ and, $N^2$ respectively, where $N$ stands for the total number of channels taken into account \cite{Manolopoulos1986}. The computational cost becomes prohibitive when the number of accessible quantum states increases, e.g., in high temperature environments when many levels can be accessed or for large molecules with small rotational constants.
Approximate techniques have been introduced to circumvent the heavy computational needs of the CC formalism, such as the Coupled States (CS) or Infinite Order Sudden (IOS) methods that are based on decoupling schemes \cite{McGuire1974}. Other approaches include the Mixed Quantum/Classical Theory (MQCT) \cite{Babikov2015,Joy2024,Bostan2024} or the Quasi-Classical Trajectory (QCT) method \cite{Faure2007_QCT,Faure2016_HC3N}. While these methods typically provide accurate results at high kinetic temperatures, their precision varies in the low temperature regime and deviations from CC rate coefficients by more than a factor 2 are not uncommon. On the other hand, statistical methods can be accurate at low temperatures \cite{Loreau2018a,Loreau2018c,Konings2021, Gonzalez-Lezana2026} but their applicability to collisions involving weakly interacting partners such as molecules colliding with H$_2$ or He is limited \cite{Loreau2018a,Balanca2020}. In addition, the errors induced by these approximate methods are not uniform and depend on the identity of the colliding molecules. 

Various attempts have been made to use ML models to improve state-to-state cross sections and rate coefficients obtained with approximate methods. 
Several of these focused on the SiO - H$_2$ system, for which it was shown that a Bayesian ML algorithm could partially correct CS cross sections using a small number of CC cross sections \cite{Jasinski2020}. Another recent work successfully used a Multi Layer Perceptron architecture to predict the difference between CS and CC rate coefficients for rovibrational transitions \cite{Wang2025}. GP regression was also employed to speed up the calculations by 75\% by only calculating a subset of all state-to-state rate coefficients \cite{Mihalik2025}. As another example, Bossion et al. employed a neural network model that utilizes a small amount of Time-Independent Quantum Mechanical (TIQM) data to optimize the QCT rate constants of the H + H$_2$ reaction, which contains both reactive and inelastic channels. It was shown that in the low temperature region, where QCT calculations perform poorly, quantum effects could be recovered to an extent by such a model.

Although the framework of $\Delta$-learning models is promising for such applications, they still require high-quality (i.e., CC) data for the systems at hand, which can remain an obstacle for collisions involving heavy molecules or proceeding through deep potential wells. To circumvent these issues, different approaches have been proposed based on models trained on only a sparse amount of accurate data to generate high quality results through interpolation. A recent example is the work conducted on the rotational transitions of H$_2$O + H$_2$O collisions \cite{Mandal2025_H2O_ML}, a system of interest for cometary applications. MQCT calculations for such collisions involving two asymmetric top molecules take several millions CPU hours, but interpolation across a fragmentary dataset with an ANN model focusing on the physical relations could yield cross sections with good accuracy. Similarly, ML approaches were explored to complete datasets of rate coefficients for vibrational energy transfer in N$_2$ + H$_2$ collisions \cite{Hong2023} or to provide pressure broadening parameters \cite{Guest2024_broadening}.

Pure data-driven approaches can also be developed to capture the underlying physics in molecular collisions, which was originally envisioned by Neufeld \cite{Neufeld2010_ANN}. Despite the limited datasets available data at the time of his study, he showed the possibility of estimating rotational excitation rate coefficients for  transitions involving highly excited levels by training an ANN with a small set of descriptors. However, the accuracy of the model was limited, with average errors larger than a factor 2 for H$_2$O-H$_2$ and SO$_2$-H$_2$ collisions. 

Two obstacles stand in the way of the use of ML methods in the prediction of rate coefficients for molecular collisions: the datasets need to be consistent (i.e., computed at the same level of theory) and large enough, since the performance of ML models usually increases with the size of the training set.
Large amounts of temperature-dependent collisional excitation rate coefficients are nowadays available in dedicated databases for about a hundred molecules \cite{EMAA, basecol2023} and mostly for rotational excitation. While these data are typically obtained through a combination of high-quality quantum chemistry (e.g., CCSD(T)) and quantum dynamics (e.g., CC or CS) approaches, they are restricted to a certain range of kinetic temperatures and, more critically, rotational levels, due to the choice of the rotational basis used to expand the wavefunction. While such data are suitable to build radiative transfer models of the cold interstellar medium, this is often not the case for warm astrophysical environments. 
This challenge is even more significant when considering the collisional excitation of complex organic molecules (COMs, defined in an astrochemistry context as having six or more atoms) that have a large number of rotational levels populated even at low temperatures \cite{Faure2011a,Dzenis2022,BenKhalifa2024a,BenKhalifa2025a}.
Expanding state-to-state rate coefficient datasets to cover a larger number of rotational levels can be done by computing high-energy cross sections with a cheaper numerical method, but combining results obtained with different dynamical approaches is non-trivial. 

In this study, we examine the potential of ANNs with physically motivated descriptors to robustly extend existing datasets of state-to-state rate coefficients for molecular collisions, thus bypassing the need for additional scattering calculations. 
Initially, a Multi Layer Perceptron (MLP) model, a specific class of ANN, is benchmarked for predicting the inelastic rate coefficients by systematically optimizing the model structure, hyperparameters and regularization parameters. We aim to test the predictive capability of the ANN model through extrapolation to higher rotational levels, higher kinetic temperatures and for both of these conditions simultaneously.
Subsequently, the model's generalizability is verified by testing it on the excitation of molecules exhibiting different characteristics in collisions with He atoms and H$_2$ molecules, for which highly accurate reference datasets of state-to-state rate coefficients are available. These include CO + H$_2$ and SiO + H$_2$ (closed shell linear molecules),  SO + H$_2$ (open-shell linear molecule), H$_2$O + H$_2$ and SO$_2$ + H$_2$ (small asymmetric top molecules), and vinyl cyanide (CH$_2$CHCN) and benzonitrile (C$_6$H$_5$CN), which provide examples of the excitation of COMs (large asymmetric top molecules) 
induced by He atoms, for which quantum scattering calculations are extremely challenging due to the high density of the rotational levels.

The paper is structured as follows: In Sec. \ref{sec:model} we discuss the architecture of the ANN model; while results are presented in Sec. \ref{sec:results}. Sec. \ref{subsec:extrap_levels} presents an analysis of the model performance for extrapolation over unseen rotational levels for two dissimilar molecules (CO and C$_6$H$_5$CN).
Sec. \ref{subsec:extrap_temperature} focuses on an extrapolation towards higher temperatures only, or in combination with an extrapolation on rotational levels. In Sec. \ref{subsec:dataset_size} we analyse the impact of the training set and test set size. Finally, in Sec. \ref{subsec:all_molecules} we examine the generality of the NN model by investigating a range of other molecular systems. Conclusions are drawn in Sec. \ref{sec:conclusions}.

\section{Machine Learning Framework}\label{sec:model}

\subsection{Artificial Neural Network}

We employ a Multilayer Perceptron (MLP) as the ANN architecture, which consists of multiple interconnected neurons organized into distinct layers, implemented in PyTorch.  The architecture of the model comprises an input layer, two hidden layers each consisting of 128 neurons, and an output layer. This choice corresponds to the ANN model providing the best performance according to tests carried out with various numbers of hidden layers ($2-4$)  each containing various amounts of neurons ($16-256$).
The input neurons are fed with the relevant collision features (see below) and this information is processed and propagated to every neuron in the subsequent layer. The operation performed for each neuron is expressed as:
\begin{equation}\label{eq_neurons}
	h_j^{(l)} = f\left( \sum_{i=1}^{N_{l-1}} w_{ji}^{(l)} h_i^{(l-1)} + b_j^{(l)} \right)
\end{equation}
where $h_j^{(l)}$ denotes the output of neuron $j$ in layer $l$, $N_{l-1}$ is the total number of neurons in the previous layer, $w_{ji}^{(l)}$ is the weight that is connecting neuron $i$ in the previous layer to neuron $j$ in the current layer, $h_i^{(l-1)}$ is the incoming output from neuron $i$ (for the first hidden layer, these correspond to the initial input features $x_i$), $b_j^{(l)}$ represents the bias term for neuron $j$, and $f$ designates the non-linear activation function.
Ultimately, all neurons in the final hidden layer were connected to a single output neuron to provide the predicted state-to-state rate coefficients. 

In our model, the Leaky Rectified Linear Unit (Leaky ReLU), a modified version of the standard RELU, was employed as the activation function for all hidden layers. Unlike the standard ReLU, it allows a small, non-zero gradient to prevent neurons from becoming permanently inactive and stopping the propagation of information:
\begin{equation}
	f(z) = \max(0, z) + \alpha \min(0, z)
\end{equation}
where $z$ is the pre-activation sum defined in equation (\ref{eq_neurons}) and $\alpha$ is a small constant that was set to $\alpha=0.01$.  By assigning a small weight to negative inputs, it ensures continuous gradient flow and maintains an active learning process.

Initial weight values were generated using the Kaiming (He) normal initialization technique to prevent the exploding or vanishing gradient problems at the beginning of the training phase. This method is specifically optimized for ReLU-based activation functions and it generates initial weights by using a normal distribution having a variance of Var$(W) = 2/N_{in}$, where $N_{in}$ represents the number of incoming neurons for that specific layer.

\subsection{Training and Optimization}

Since state-to-state collisional excitation rate coefficients ($k(T)$) span values across several orders of magnitude, training the model directly on these values is a substantial challenge. Rather, the training is realized by using the logarithm of the rate coefficients, $y = \log_{10}(k(T))$.

The Mean Squared Error (MSE) loss function was utilized for optimization to train the network within this logarithmic space. Furthermore, an $L_2$ regularization penalty (weight decay) was included to prevent the model from overfitting the data and to restrict the uncontrolled growth of weight values. Consequently, the total loss function ($\mathcal{L}$) that the model aimed to minimize is expressed as 
\begin{equation}
	\mathcal{L} = \frac{1}{N} \sum_{i=1}^{N} (y_i - \hat{y}_i)^2 + \lambda \sum_{l} \|\mathbf{W}^{(l)}\|_2^2
\end{equation}
where $y_i$ is the actual logarithmic rate coefficient, $\hat{y}_i$ is the predicted logarithmic rate coefficient, $\mathbf{W}^{(l)}$ denotes the weight matrices across all layers, and $\lambda$ represents the regularization coefficient which was set to $10^{-4}$ for our model.

The Adam (Adaptive Moment Estimation) Optimizer was employed to minimize the complex multidimensional loss landscape. Unlike standard stochastic gradient descent, the learning rates are updated in an individual and adaptive manner for each parameter. The Adam optimizer takes information about preceding gradients into account to make calculations more efficient. Exponential moving averages of the first (mean) and second (variance) moments of the gradients of the loss function are utilized for this procedure. At any given step $t$, these moments are calculated as:
\begin{equation}
	m_t = \beta_1 m_{t-1} + (1 - \beta_1) g_t
\end{equation}
\begin{equation}
	v_t = \beta_2 v_{t-1} + (1 - \beta_2) g_t^2
\end{equation}
where $g_t$ is the gradient of the loss function and $m_t$ and $v_t$ represent the first and second moments, respectively. The damping coefficients were set to $\beta_1 = 0.9$ and $\beta_2 = 0.999$. Since $m_t$ and $v_t$ are initialized as 0 at the beginning of training, a bias correction, which affects the initial moments but vanishes in subsequent steps, was implemented to circumvent the tendency of these moments to bias towards zero:
\begin{equation}
	\hat{m}_t = \frac{m_t}{1 - \beta_1^t}
\end{equation}
\begin{equation}
	\quad \hat{v}_t = \frac{v_t}{1 - \beta_2^t}
\end{equation}
The entire set of parameters $\theta$ encompassing the values of weights and biases was updated as 
\begin{equation}
	\theta_t = \theta_{t-1} - \frac{\eta_t}{\sqrt{\hat{v}_t} + \epsilon} \hat{m}_t
\end{equation}
where $\epsilon=10^{-8}$ and $\eta_t$ is the learning rate. 
To ensure a smooth convergence, a learning rate scheduler (LambdaLR) was included in the model. It works as an inverse time decay function that gradually reduces the initial learning rate as the number of iterations increases:
\begin{equation}
	\eta_t = \eta_0 \left( \frac{1}{1 + c \cdot t} \right)
\end{equation}
where $\eta_0=0.01$ is the initial learning rate, $t$ is the current epoch and $c=10^{-4}$ is a damping coefficient.

As the training phase extends, neural networks can start to memorize the given training data instead of learning the underlying physics of the system. This situation degrades the prediction performance of the model on unseen scenarios. To prevent such overfitting, an early stopping mechanism was employed.
In every epoch, the loss function $\mathcal{L}_{val}$ was evaluated on the validation set. The algorithm checked at every step whether the current $\mathcal{L}_{val}$ was smaller than the previous minimum. When a smaller value was encountered, all weights and biases at that epoch were saved for $\mathcal{L}_{best}$. If a smaller value was not encountered within a patience limit of 1000 epochs, the model returned to the weights and biases for $\mathcal{L}_{best}$ and training was terminated.

\subsection{Evaluation Metrics}

Assessments of the model's predictive power were carried out with different error metrics.
In this section, the actual value of the rate coefficients are denoted by $k$, predicted rate coefficients are identified as $\hat{k}$ and their logarithmic correspondences are denoted by $y = \log_{10}(k)$ and  $\hat{y} = \log_{10}(\hat{k})$, respectively.

Traditionally, the most widely used error metric is the Mean Average Percentage Error (MAPE),
\begin{equation}
\text{MAPE} = \frac{1}{N} \sum_{i=1}^{N} \left| \frac{k_i - \hat{k}_i}{k_i} \right| \times 100\% \ .
\end{equation}
However, since the rate coefficients span several orders of magnitude in a datatset, a prediction that is wrong by even one order of magnitude results in a meaningless MAPE value. Moreover, the relative differences between actual and predicted values lead to an asymmetry problem. For example,  evaluating the error with $k_i = 10^{-15}$ cm$^{-3}$ and $\hat{k}_i = 10^{-12}$ cm$^{-3}$ yields a drastically different result compared to the reverse scenario with $k_i = 10^{-12}$ cm$^{-3}$ and $\hat{k}_i = 10^{-15}$ cm$^{-3}$. While this can be circumvented through the use of the symmetric Mean Average Percentage Error (sMAPE), a simpler, more robust, and more intuitive measure is the Mean Factor Error (MEF):

\begin{equation}
\text{MEF} = \frac{1}{N} \sum_{i=1}^{N} 10^{\vert y_i - \hat{y}_i \vert}
\end{equation}
This definition ensures that an overestimation or an underestimation of a rate coefficient by a given factor corresponds to the same MEF. 

Another useful error measure is the Weighted Mean Factor Error (WMEF) \cite{Loreau2018a},
\begin{equation}
\text{WMEF} = \frac{ \sum_{i=1}^{N} k_i 10^{|y_i - \hat{y}_i|} } {\sum_{i=1}^{N} k_i}
\end{equation}
which gives more weight to larger rate coefficients. This reflects the fact that when using inelastic collisions rate coefficients datasets in  non-local thermodynamic equilibrium radiative transfer models for astrochemical applications, the dominant rate coefficients play a larger role, so that they should be predicted with a better accuracy. 

Finally, the Root Mean Square Error (RMSE) was also used, as it is more sensitive to outliers. We define it with respect to the $\log_{10}$ of the rate coefficients, 
\begin{equation}
\text{RMSE} = \sqrt{\frac{1}{N} \sum_{i=1}^{N} (y_i - \hat{y}_i)^2}
\end{equation}

\subsection{Training data and input features}\label{subsec:input_features}

The full datasets of state-to-state rate coefficients $k_{ij}(T)$ for initial and final quantum state $i$ and $j$ were downloaded from the EMAA \cite{EMAA} and BASECOL \cite{basecol2023} databases unless stated otherwise. The datasets contain only de-excitation rate coefficients, as the excitation rate coefficients can be re-constructed from detailed balance. Moreover, de-excitation rate coefficients advantageously cover a smaller range of values than excitation rate coefficients.
The datasets are split into a set that includes rate coefficients for transitions involving all quantum states up to a fixed value and/or a fixed temperature. 90\% of these data are randomly chosen to use as the training  set, while the remaining 10\% form the validation set. Finally, state-to-state rate coefficients involving higher-lying quantum states and/or temperatures form the test set.

The input features include the kinetic temperature ($T$ in kelvin) and the ratio $\Delta E / k_B T$, where $\Delta E = E_i - E_f$ is the energy difference between the initial and final quantum state. The latter follows from the expectation that inelastic rate coefficients conform to the energy gap law for large energy gaps. Moreover, we also use the energy of the initial and final level ($E_i$ and $E_f$) as this was found to improve the performance of the model.
Finally, the values of the relevant quantum numbers of the initial and final states of the colliding molecules as well as the absolute value of their differences are fed into the model, which is crucial for learning the propensity rules that determine which transitions are more likely to occur in molecular collisions.

The various input features fed into the model possess different scales. Without any standardization, the features with large absolute values would dominate the gradients of the network. 
To prevent this issue, Z-score normalization was applied to all features independently:
\begin{equation}
	z_i^{(j)} = \frac{x_i^{(j)} - \mu^{(j)}}{\sigma^{(j)}}
\end{equation}
where $x_i^{(j)}$ represents the raw value of the $i^{\textrm{th}}$ data point of the type $j$ feature, $\mu^{(j)}$ is the mean of this feature and $\sigma^{(j)}$ is the corresponding standard deviation.
To prevent data leakage, the standardization process was conducted using solely the training data. The mean and standard deviation parameters were calculated exclusively from the training set and subsequently applied to normalize the validation and test sets.


\section{Results and discussion}\label{sec:results}

We first examine the performance of our model by considering two datasets that correspond to vastly different types of inelastic molecular collisions, namely the excitation of carbon monoxide (CO) by \textit{para}-H$_2$ in its ground rotational level and of benzonitrile (C$_6$H$_5$CN) by He atoms. We focus in both cases on the pure rotational excitation with datasets containing only de-excitation rate coefficients. CO is ubiquitous in astrophysical environments where it can be used as a probe of the physical conditions. Such a diatomic molecule is described by a single quantum number $j$ and displays a simple rotational structure, which an ANN model could be expected to be able to describe adequately. On the other hand the dataset \cite{Yang2010} for CO-\textit{para}-H$_2$ (hereafter simply denoted as CO-H$_2$) collisions covers a wide range of temperatures (from 2 K up to 3000 K) and the corresponding rate coefficients vary by 13 orders of magnitude (from approximately 10$^{-23}$ cm$^3$s$^{-1}$ to 10$^{-10}$ cm$^3$s$^{-1}$), which might be difficult to capture with a single model. The full dataset covers transitions between the lowest 41 rotational levels (up to $j=40$), or 820 rate coefficients for each temperature, and 25 values of the temperature. The inelastic rate coefficients were computed based on scattering calculations with the CC method for collision energies up to 800~cm$^{-1}$ and with the CS technique for higher energies \cite{Yang2010}. We note that other datasets are available for CO-H$_2$ collisions \cite{Faure2006_CO,Dagdigian2022_CO} but these cover a smaller number of rotational levels.

On the other hand, benzonitrile is a complex asymmetric top molecule with a rotational structure defined by three quantum numbers $j$, $k_a$, and $k_c$. It has been detected in a variety of astrophysical environments \cite{mcguire2018_benzo,Burkhardt2021_benzo} where it is considered a key species to understand the formation of cyclic and polycyclic molecules.
Due to its large molecular mass, the rotational level manifold is extremely dense, and with 13 atoms, benzonitrile is the largest molecule for which a dataset of inelastic rate coefficients is available \cite{BenKhalifa2024a}. In addition, due to the symmetry of the hydrogen atoms, benzonitrile possesses \textit{ortho} and \textit{para} forms (for $k_a$ even or odd, respectively) and the interconversion between these two forms is forbidden through inelastic collisions. 
The dataset covers transitions between the lowest 110 rotational levels of \textit{para}-C$_6$H$_5$CN and 104 rotational levels of \textit{ortho}-C$_6$H$_5$CN induced by collisions with He atoms, which are often used as proxies for ground state \textit{para}-H$_2$ in scattering studies. The inelastic rate coefficients were computed by means of the coupled states method and were shown to agree well with CC calculations, as detailed in Ref. \cite{BenKhalifa2024a}. The full dataset contains 5356 rate coefficients for \textit{ortho}-C$_6$H$_5$CN and 5995 rate coefficients for \textit{para}-C$_6$H$_5$CN, scaled to account for the reduced mass difference between He and H$_2$, at 11 temperatures between 5 and 100 K, and was used recently to study non-LTE effects in benzonitrile observations \cite{Rivilla2026_benzo}. 
While the rotational structure of benzonitrile is much more complex than that of CO, the rate coefficients cover a smaller temperature range and display less variation, with all rate coefficients having values between 10$^{-15}$ cm$^{-3}$ and 10$^{-10}$ cm$^{-3}$. This can be expected to improve the predictive capability of the NN model. 
For CO the rotational quantum number of the initial and final states, $j_i$ and $j_f$, as well as $\vert j_f - j_i \vert$, are used as input features (see Sec. \ref{subsec:input_features}), while for C$_6$H$_5$CN we use the initial and final values of $j$, $k_a$, $k_c$, as well as the absolute value of their difference.

\subsection{Extrapolating towards high rotational levels}\label{subsec:extrap_levels}

For both molecules, the model detailed in Section \ref{sec:model} was used to predict inelastic rate coefficients for transitions involving the highest 20\% of all rotational levels at all temperatures.
 
Specifically, for CO-H$_2$ the training set contains 90\% of all transitions between the first 33 levels ($j=0-32$, with 528 rate coefficients per temperature and 13200 data points in total) while the remaining 10\% is used as a validation set. The test set contains all de-excitation rate coefficients with initial levels $34-41$ ($j=33-40$) and final levels 1-40 ($j=0-39$), with 292 rate coefficients per temperature. The size of the test set is thus 62\% of the size of the training set.
For C$_6$H$_5$CN-He a similar division is applied and the training set contains 90\% of all transitions between the first 87 levels for \textit{para}-C$_6$H$_5$CN and 82 levels for \textit{ortho}-C$_6$H$_5$CN (3741 and 3321 rate coefficients per temperature respectively) while the test set contains all de-excitation rate coefficients starting from levels $88-110$ for \textit{para}-C$_6$H$_5$CN and $83-104$ for \textit{ortho}-C$_6$H$_5$CN (2254 and 2035 rate coefficients per temperature respectively). The size of the test set is thus approximately 67\% of the size of the training set in both cases. 
The impact of the size of the training and test sets is further examined in Section \ref{subsec:dataset_size}. Since the results were found to be qualitatively similar for both spin isomers of benzonitrile, we will focus our analysis solely on \textit{para}-C$_6$H$_5$CN. The list of levels of \textit{para}-C$_6$H$_5$CN along with their associated quantum numbers and energies is given in Table \ref{table_levels_pBN} of the Appendix.

The first test of the model performance is the error on the validation set, which contains transitions involving the same rotational levels included in the training set. The MEF on the validation set is 1.07 for CO and 1.09 for \textit{p}-C$_6$H$_5$CN, while the corresponding RMSE values (on $\log_{10}{k}$) are 0.0410 and 0.0567, respectively. These low errors demonstrate that the model is able to learn the main trends in the rate coefficients over the whole range of temperature for both molecules and interpolate the missing data, as has been shown in previous works \cite{Mihalik2025,Mandal2025_H2O_ML}. The slightly larger value for C$_6$H$_5$CN is likely a reflection of  the fact that the propensities in the rate coefficients are more complex for asymmetric tops than for closed shell linear molecules. 

\begin{figure}
\center
\includegraphics*[width=.9\linewidth]{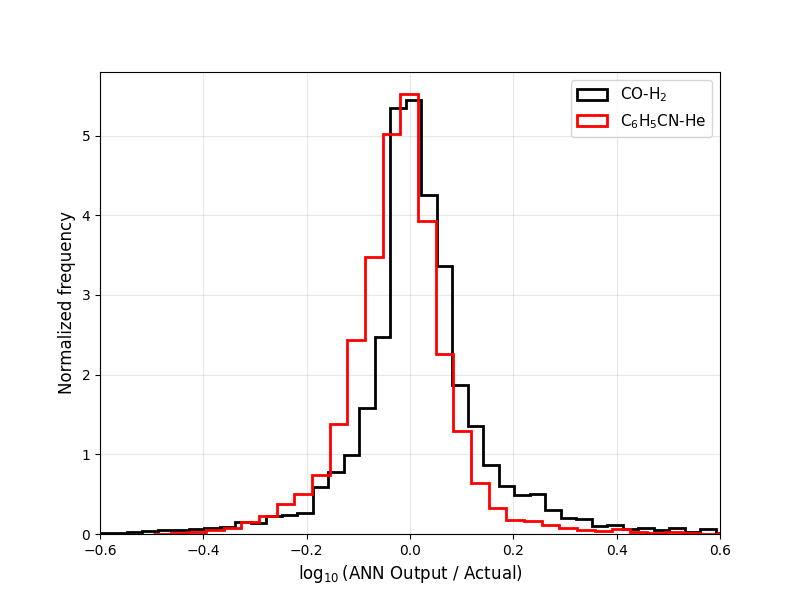}
\caption{Frequency distribution of errors (in log scale) of predicted rate coefficients for CO-H$_2$ (transitions involving initial levels 34-41) and \textit{p}-C$_6$H$_5$CN-He (transitions involving initial levels 88-110).}
\label{fig:histogram_CO_BN}
\end{figure}

Figure \ref{fig:histogram_CO_BN} displays the frequency distribution of errors (in log scale) in the prediction of CO-H$_2$ rate coefficients for initial levels $34-41$ and \textit{p}-C$_6$H$_5$CN-He rate coefficients for initial levels $88-110$ across all temperatures. In both cases the distribution is centered around 0 and close to symmetric with median values of 0.01 (see also Table \ref{table_errors_CO_pBN}), indicating that the prediction of unknown rate coefficients is not systematically biased towards values that are either too large or too small. 

\begin{figure}
\center
\includegraphics*[width=.9\linewidth]{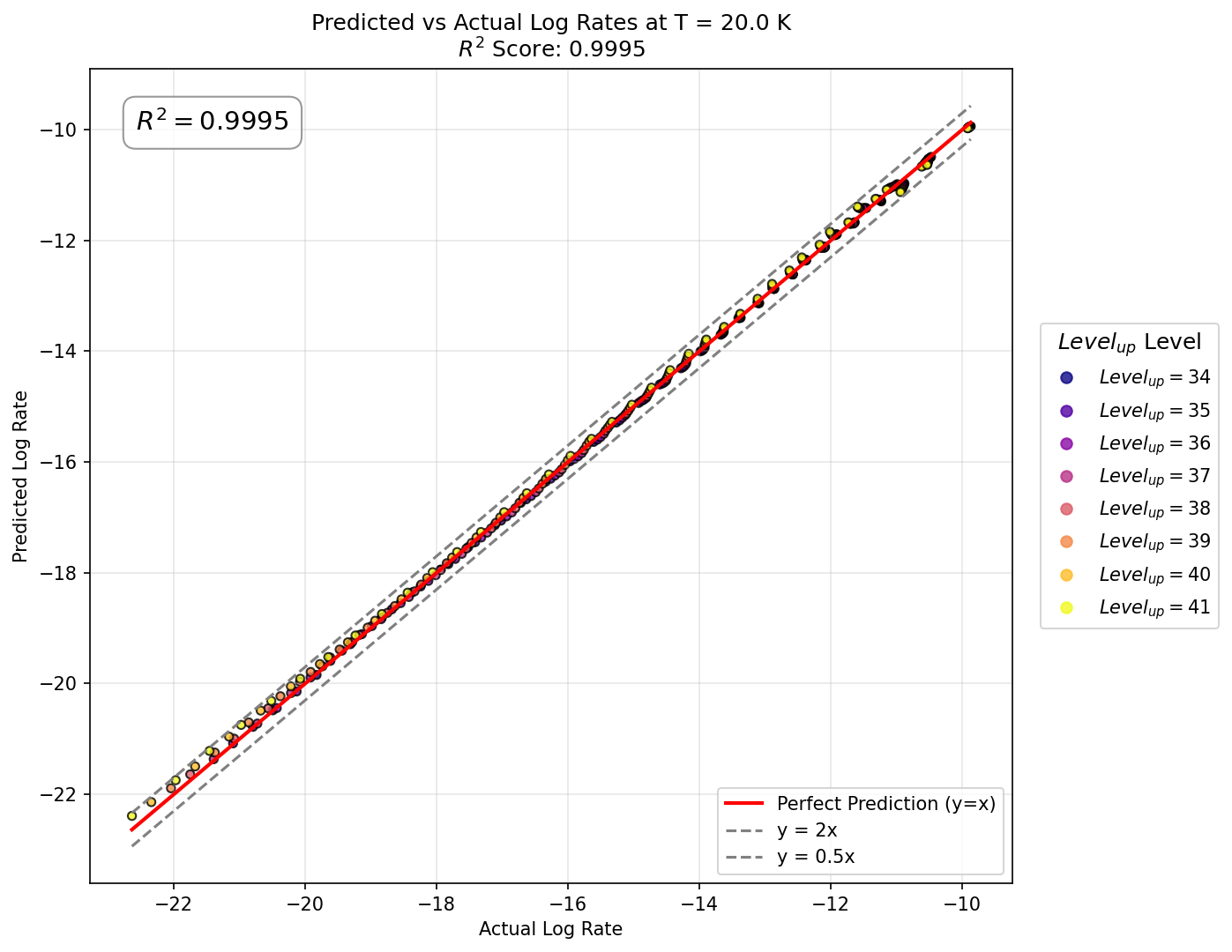}
\caption{Scatter plot of predicted versus actual CO-H$_2$ rate coefficients for transitions involving initial levels 34-41 at a kinetic temperature of 20 K.}
\label{fig:CO_scatter}
\end{figure}

\begin{figure}
\center
\includegraphics*[width=.9\linewidth]{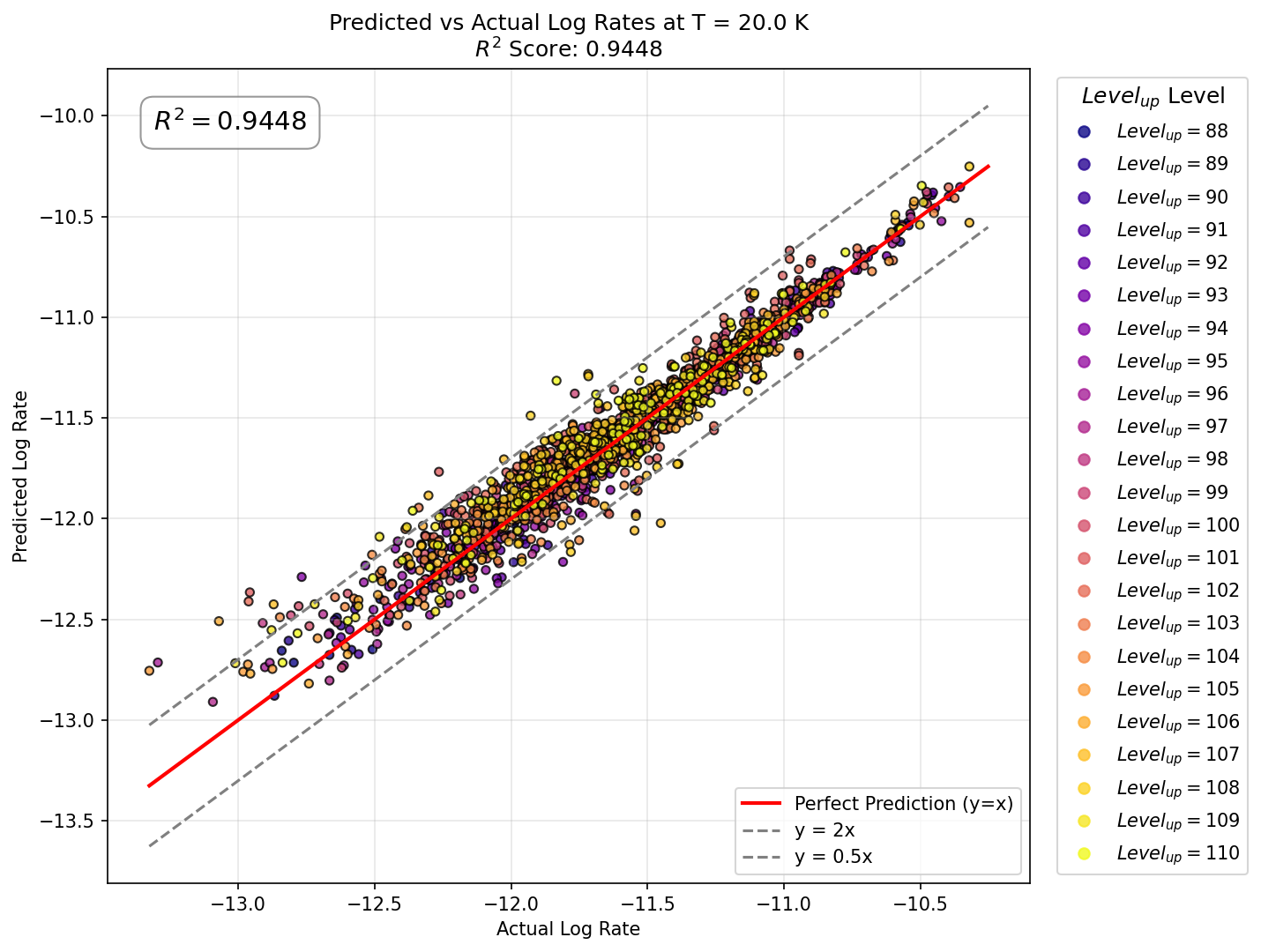}
\caption{Scatter plot of predicted versus actual \textit{p}-C$_6$H$_5$CN-He rate coefficients for transitions involving initial levels 88-110 at a kinetic temperature of 20 K.}
\label{fig:pBN_scatter}
\end{figure}

Fig. \ref{fig:CO_scatter} compares the model prediction to the actual value of the rate coefficients for CO-H$_2$ at a fixed temperature of 20 K. A similar comparison is displayed in Figs. S1-S2 of the Supplementary Material for temperatures of 100 K and 500 K. We observe all rate coefficients are reproduced to better than a factor 2, with a MEF and WMEF of 1.13 and 1.15 respectively. The MEF and WMEF averaged over all temperatures are only 1.27 and 1.17 respectively (see Table \ref{table_errors_CO_pBN}) despite the wide range of values of the inelastic rate coefficients, while the RMSE value is 0.109.
The lower value of the WMEF compared to MEF indicates that the error is smaller for the larger rate coefficients, which is a desirable outcome. 
The coefficient of determination is $R^2=0.9995$, confirming that the rate coefficients are nearly perfectly extrapolated.
Fig. \ref{fig:pBN_scatter} similarly compares the model prediction to the actual value of the rate coefficients for \textit{p}-C$_6$H$_5$CN at the same fixed temperature of 20 K. In this case almost all rate coefficients are reproduced to better than a factor 2, although a small fraction are outside this range. 
The MEF and WMEF of 1.13 and 1.15, respectively, at 20 K, while the values averaged over the whole temperature range result in a MEF of 1.22 and a WMEF of 1.17 (see Table \ref{table_errors_CO_pBN}). These values are smaller or equal to those of CO-H$_2$ despite the increased molecular complexity. This observation, which might be unexpected, can be rationalized by the fact that the size of the training dataset is much larger for \textit{p}-C$_6$H$_5$CN, which improves the accuracy of the model. In fact, we observed that increasing the size of the training set by including data for the excitation of both the \textit{o}-C$_6$H$_5$CN and \textit{p}-C$_6$H$_5$CN spin isomers also improved the quality of the predictions, despite the fact that they can be considered as independent species in inelastic collisions.
In addition, the temperature range is much smaller for \textit{p}-C$_6$H$_5$CN-He (5-100 K, compared to 2-3000 K CO-H$_2$) and the values of rate coefficients cover a smaller range. The overall RMSE is 0.08, which is again smaller than the RMSE obtained for CO-H$_2$. The coefficient of determination is $R^2=0.9448$, indicating more variance in the prediction than for the case of CO-H$_2$. 


\begin{figure}
\includegraphics*[width=.9\linewidth]{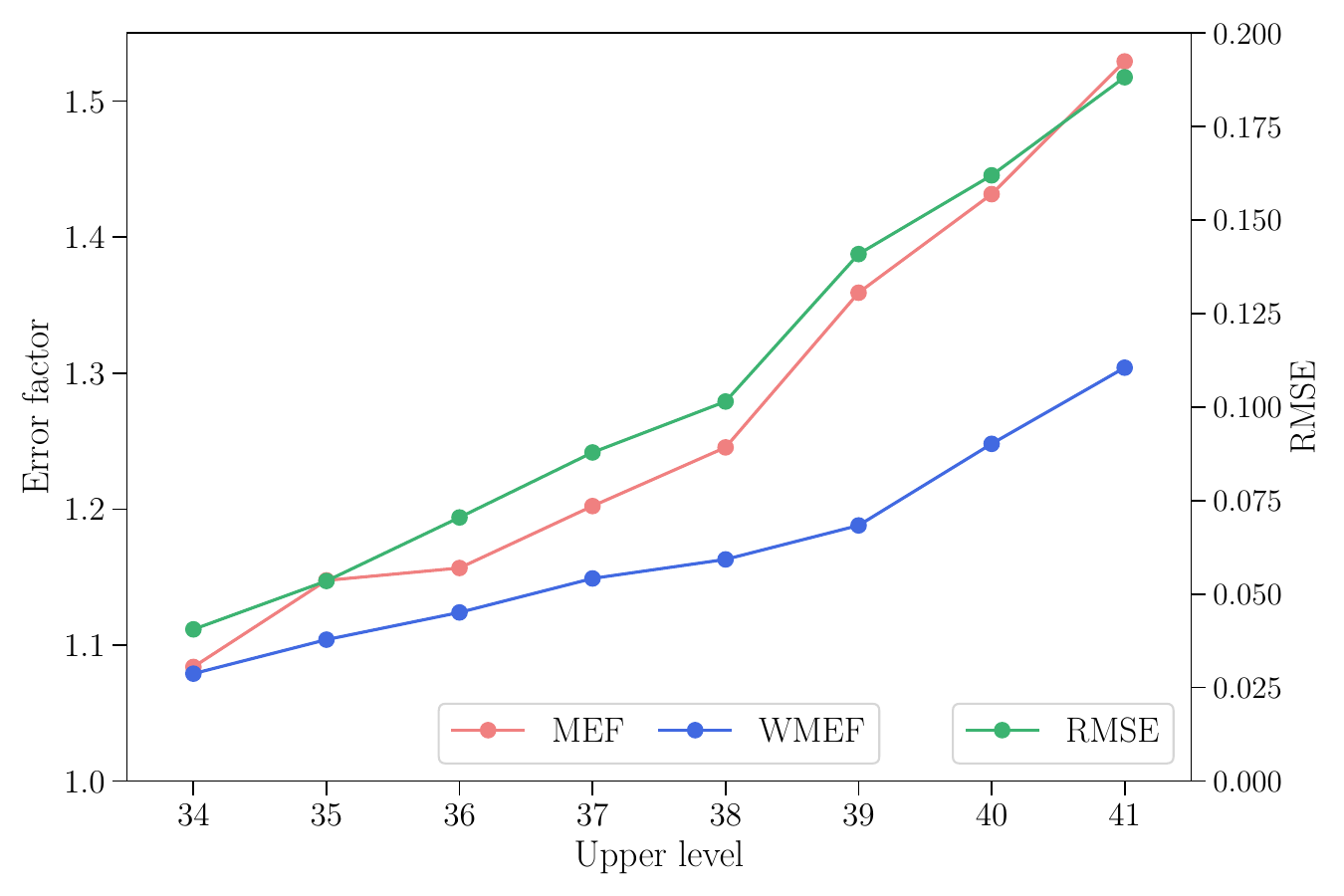}
\caption{Error measures (MEF and WMEF, left scale; RMSE, right scale) of predicted CO-H$_2$ rate coefficients for transitions involving initial levels 34-41, as function of the initial level and averaged over all final levels and temperatures.}
\label{fig:CO_error_level}
\end{figure}

\begin{figure}
\includegraphics*[width=.9\linewidth]{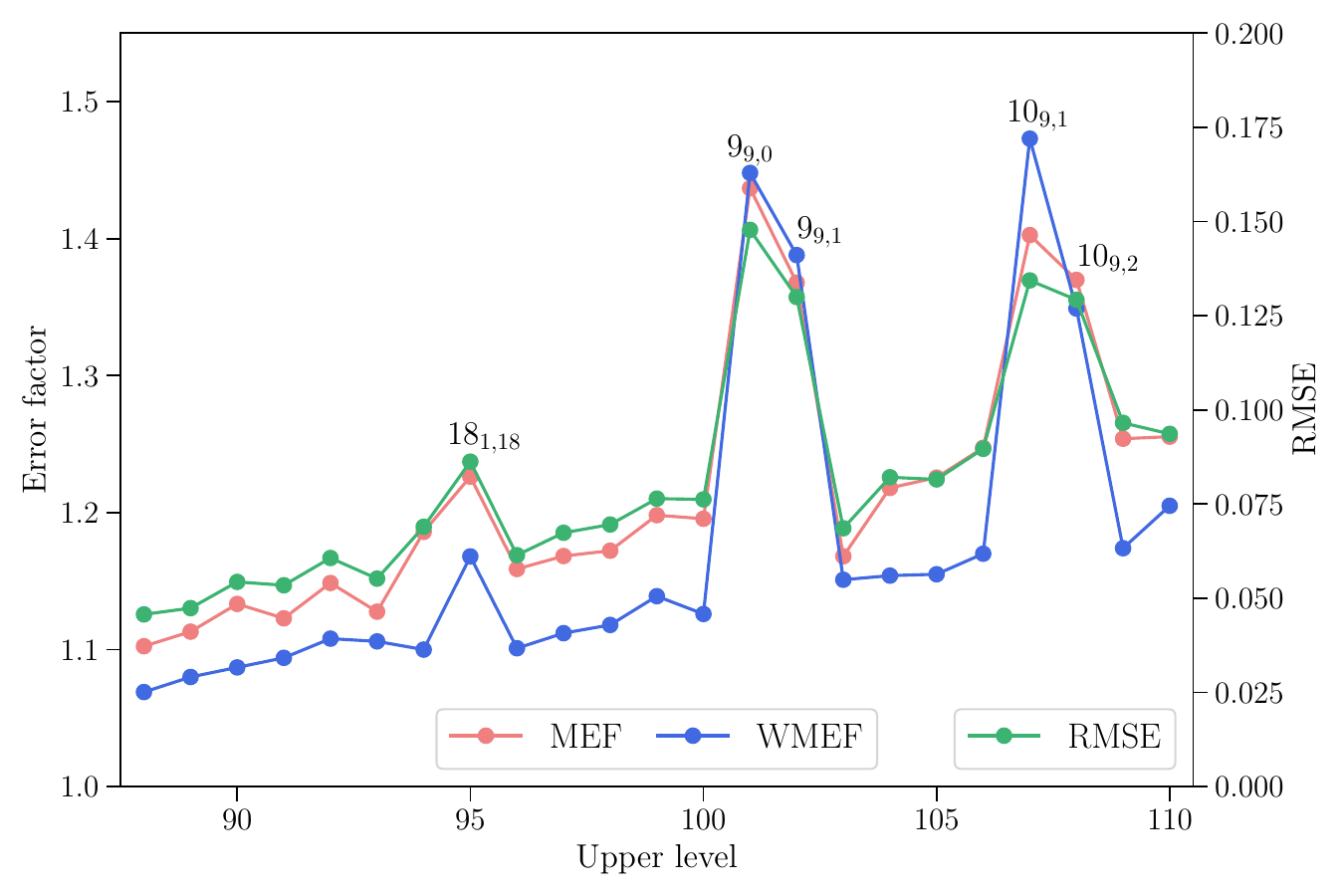}
\caption{Error measures (MEF and WMEF, left scale; RMSE, right scale) of predicted \textit{p}-C$_6$H$_5$CN-He rate coefficients for transitions involving initial levels 88-110, as function of the initial level and averaged over all final levels and temperatures. The rotational quantum numbers $j_{k_a,k_c}$ are indicated for the initial levels with the largest error (see text).}
\label{fig:pBN_error_level}
\end{figure}

The various error measures are displayed as function of the initial extrapolated level (and averaged over all final levels) in Fig. \ref{fig:CO_error_level} for CO-H$_2$. The trend for CO is evident, with a monotonous increase of all error measures with increasing initial level. Since the levels are characterized by a single rotational quantum number $j$, this corresponds to the expectation of an increasing error as the value taken by the descriptors (here $j$) moves further and further away from the values that the model was trained on. Despite this, it is clear that the model is able to properly extrapolate with a MEF factor for the highest initial rotational level ($j=41$) being of 1.5, corresponding to an average error of 50\% on average for all de-excitation rate coefficients starting from this level. 
The value of the WMEF is systematically lower than the MEF, again demonstrating that the error is smaller for the larger rate coefficients, as can be also be seen on Fig. \ref{fig:CO_scatter}.
Fig. \ref{fig:pBN_error_level} exhibits the error measures as function of the initial extrapolated level for \textit{p}-C$_6$H$_5$CN-He collisions. An overall increase with increasing level energy is again seen, although there are obvious outliers with high errors. As indicated on the figure, these correspond to specific values of the quantum numbers that are not included in the training set: level 95 corresponds to $j_{k_a k_c}=18_{1,18}$, levels 101 and 102 to levels $9_{9,0}$ and $9_{9,1}$ respectively, while 107 and 108 can be identified as $10_{9,1}$ and $10_{9,2}$ respectively (see Table \ref{table_levels_pBN}), while the largest values of $j$, $k_a$, and $k_c$ include in the training set are 17, 8, and 17, respectively. Once again, the error increases as we extrapolate further away from the descriptors in the training set, but the errors remain with a factor 1.5, which should be acceptable for applications to astrochemistry.


\begin{figure}
\center
\includegraphics*[width=.9\linewidth]{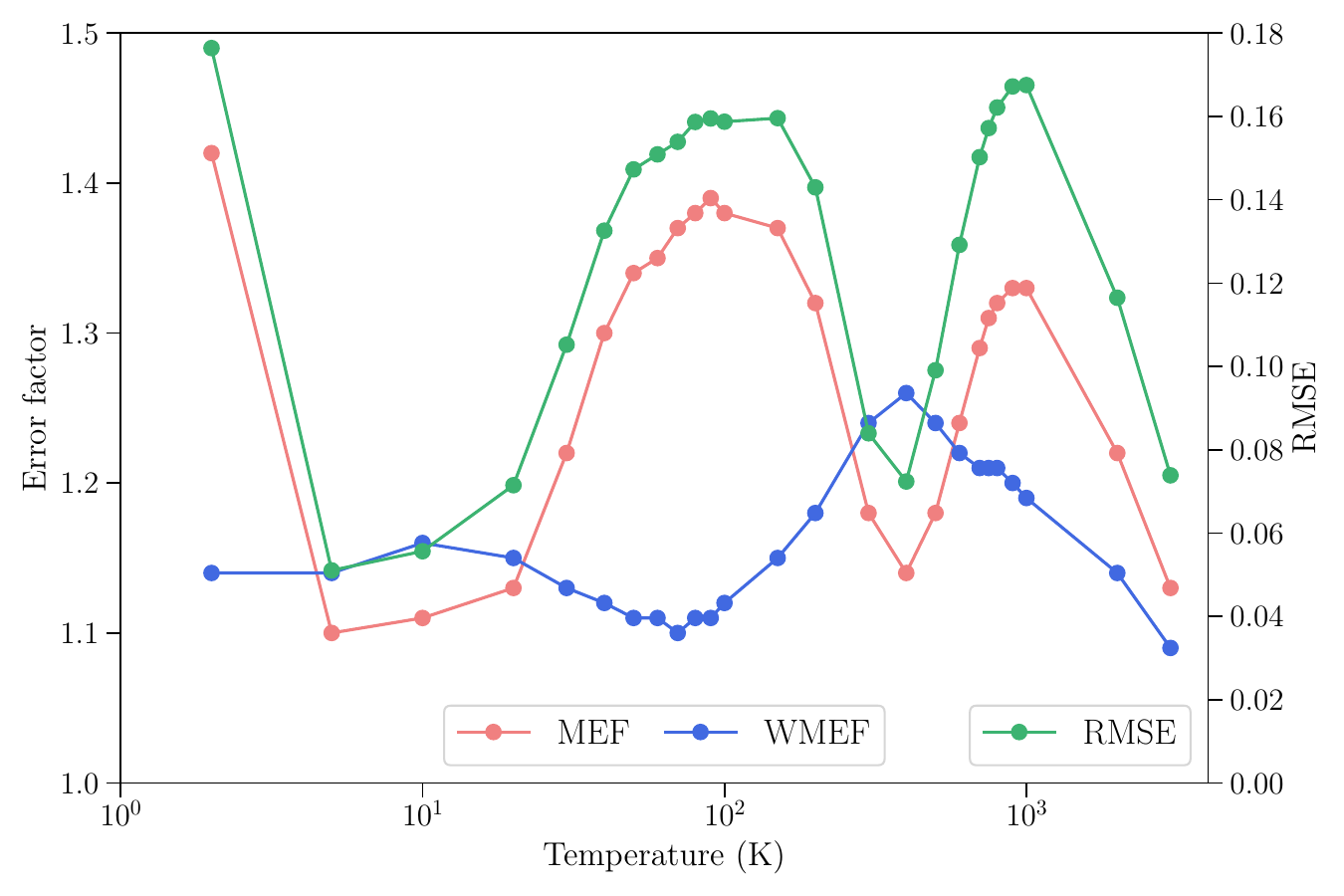}
\caption{Error measures (MEF and WMEF, left scale; RMSE, right scale) of predicted CO-H$_2$ rate coefficients for transitions involving initial levels 34-41, as function of the temperature and averaged over all initial and final levels.}
\label{fig:CO_error_temperature}
\end{figure}

\begin{figure}
\center
\includegraphics*[width=.9\linewidth]{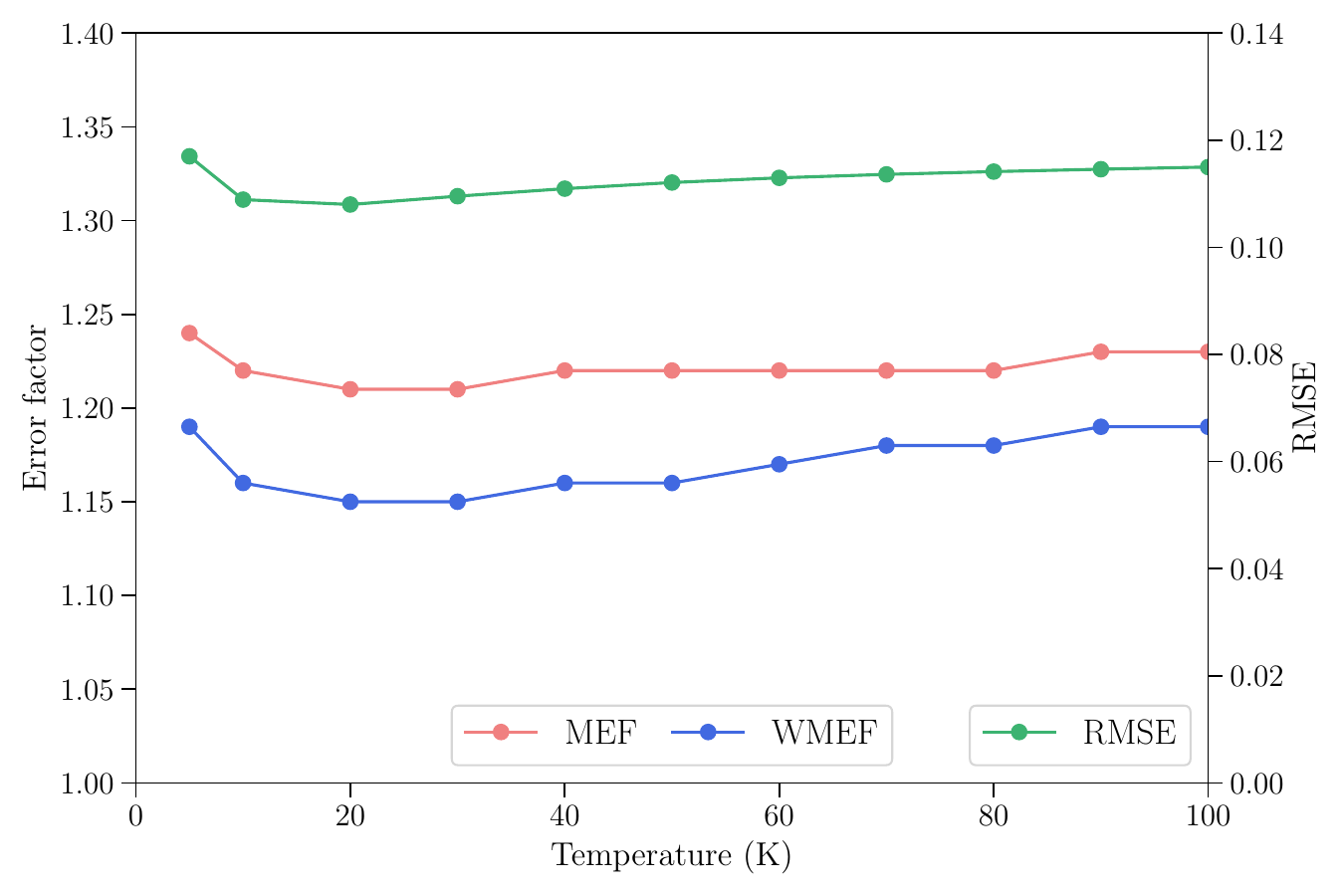}
\caption{Error measures (MEF and WMEF, left scale; RMSE, right scale) of predicted \textit{p}-C$_6$H$_5$CN-He rate coefficients for transitions involving initial levels 88-110, as function of the temperature and averaged over all initial levels.}
\label{fig:pBN_error_temperature}
\end{figure}

We next examine the behaviour of the error as function of the temperature in Fig. \ref{fig:CO_error_temperature} for CO-H$_2$ and Fig. \ref{fig:pBN_error_temperature} for \textit{p}-C$_6$H$_5$CN-He by averaging over all initial states for each temperature. In the latter case the error distribution is mostly flat, which reflects the fact that the rate coefficients themselves depend only weakly on the temperature in the range considered here \cite{BenKhalifa2024a}. For CO the situation is completely different, with large fluctuations in the MEF and RMSE values while the WMEF values depend less on temperature.
This can be explained by a combination of factors. For temperatures around 100 K and 1000 K, a relatively large error is observed on the smallest rate coefficients while the larger rate coefficients are perfectly reproduced. On the other hand, at the lowest temperature (2 K)  quantum effects such as resonances can impact the values of the rate coefficients, making model predictions more uncertain.

These results show that training the model on a large range of temperatures, which should be beneficial as it increases the size of the dataset fed to the model, might actually lead to the opposite effect due to the large range of values taken by the rate coefficients as well as possible quantum effects. As an example, retraining the model excluding the temperatures below 20 K led to markedly improved error measures (MEF of 1.16, WMEF of 1.12, RMSE of 0.09; see also Fig. S3 of the Supplementary Material). Depending on the application, it might thus be more appropriate to work with a more limited range of temperatures if possible.


\begin{table}[h!]
\centering
\footnotesize{
{\setstretch{1.5}
\begin{tabular}{c|cccc}
Molecule				& RMSE ($\log_{10}$)	& Median error ($\log_{10}$)	& MEF	& WMEF		\\ \hline
CO					& 0.109			& 0.007					& 1.27	& 1.17		\\
\textit{p}-C$_6$H$_5$CN	& 0.083			& 0.011					& 1.22	& 1.17		
\end{tabular}
}}
\caption{Overall error measures of predicted CO-\textit{p}-H$_2$ rate coefficients (initial levels 34-41) and \textit{p}-C$_6$H$_5$CN-He rate coefficients (initial levels 88-110) averaged over all initial levels and temperature.}
\label{table_errors_CO_pBN}
\end{table}

\subsection{Extrapolating towards high temperatures}\label{subsec:extrap_temperature}

A second application of the NN model is to extrapolate towards higher temperatures in order to provide rate coefficients that can be used to study non-LTE effects in warmer astrophysical environments. Given the performance of the model discussed in the previous section combined with the fact that inelastic rate coefficients usually have a smooth temperature dependence at high temperatures, such effects are expected to be captured accurately by the model.
We focus again on CO-H$_2$ and C$_6$H$_5$CN-He collisions as illustrative examples and test the approach on the extrapolation to twice the maximum temperature contained in the training set. For this purpose, we split the full CO-H$_2$ dataset into a ``low'' temperature set ($T \leq 500$ K) and a ``high'' temperature set ($T > 500$ K). 90\% of of the low temperature data is used as the training set, while the remaining 10\% of is contained in the validation set. The higher temperature rate coefficients ($500 < T \leq 1000$ K) are used as the test set.

A natural extension is to consider a simultaneous extrapolation over both the temperature and the number of rotational levels. In warmer environments a larger range of rotational levels is populated, so that generating high temperature rate coefficients without extending the dataset of rate coefficients to higher rotational levels is only of limited applicability. A third application of the ANN model is thus to train a model on the combination of the previous two types of extrapolation. In this case, the training set for CO-\textit{p}-H$_2$ includes levels 1-32 up to a temperature of 500 K, and a prediction is carried out for levels 33-41 and temperatures above 500 K.  

\begin{figure}
\center
\includegraphics*[width=.9\linewidth]{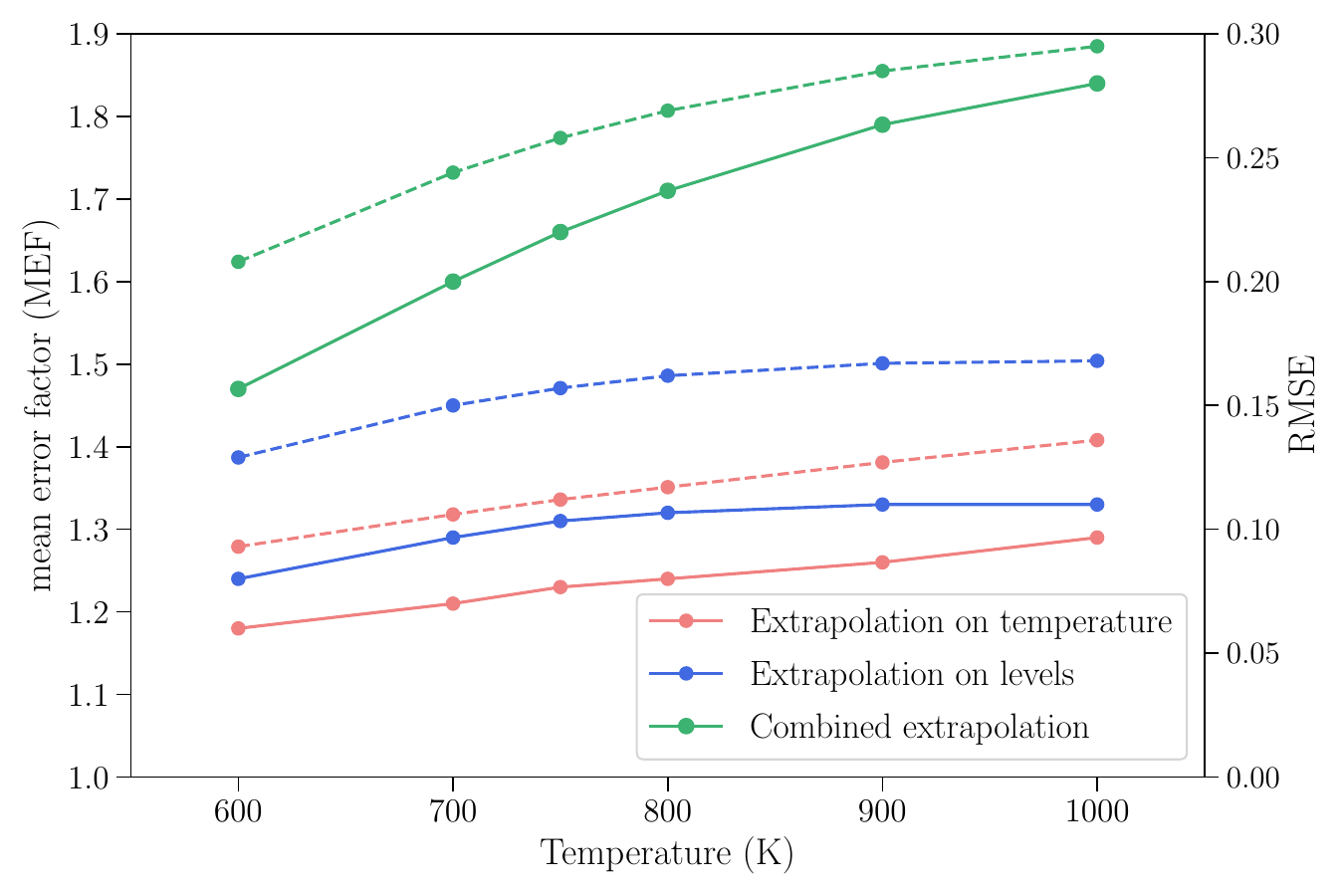}
\caption{Error measures (MEF, full lines and left scale; RMSE, dashed lines and right scale) for CO-\textit{p}-H$_2$ at high temperature for three cases: (i) extrapolation on temperatures above 500 K based on a model trained on all transitions involving levels 1-41 (red lines); (ii) extrapolation on levels 33-41 based on a model trained on all transitions involving levels 1-32 at temperatures up to 1000 K (blue lines); and (iii) cross extrapolation on temperatures above 500 K and on levels 33-41 based on a model trained on all transitions involving levels 1-32 at temperatures up to 500 K only (green lines).}
\label{fig:CO_cross_extrap}
\end{figure}

The performance of the latter two approaches can be compared to the model discussed in the previous section, which is trained on all temperatures and extrapolated to the highest rotational levels (levels 33-41) only. The comparison of the MEF and RMSE is illustrated in Fig. \ref{fig:CO_cross_extrap} and Table \ref{table_cross_extrap_CO} for temperatures above 500 K. We observe that the extrapolation only on temperatures systematically leads to the smallest errors, both in terms of MEF and RMSE. The extrapolation over rotational levels is slightly more inaccurate, but errors are still within a factor of only 1.3. These RMSE and MEF values correspond to the case examined in the previous section and illustrated in Fig. \ref{fig:CO_error_temperature}. Finally, the combined extrapolation leads to a larger error (MEF in the range 1.4 to 1.9, depending on the temperature). In both cases where an extrapolation over temperature was attempted, the error increases with temperature as we move further away from the range of the training data. It is however worth noting that despite the inherent difficulty of extrapolating over both temperatures and rotational levels, the error on the rate coefficients is still smaller then a factor 2, which is comparable to many approximate scattering methods. 
In the case of the combined extrapolation, the values of the WMEF (not shown) are lower than the MEF values, indicating that the error is smaller for the dominant rate coefficients, as in the case of the extrapolation on levels only. In addition, removing the low temperature range (< 20 K) leads to improved error measures, as was observed in the previous section for the case of extrapolation on rotational levels only.

A similar comparison of the model performance for extrapolations towards higher temperature, rotational levels, or the combination of both, was also carried out for the \textit{p}-C$_6$H$_5$CN-He rate coefficient dataset. In this case the temperature ranges from 5 K to 100 K, and the extrapolation is performed on temperatures above 50 K. The results are given in Table \ref{table_cross_extrap_pBN}. The conclusions are qualitatively similar to those found for CO-H$_2$. It is however worth noticing that the RMSE and MEF are consistently lower for \textit{p}-C$_6$H$_5$CN-He. The modest errors found for the combined extrapolation case are particularly compelling since they are comparable to the case where the extrapolation is carried out only on the rotational levels. This can be attributed to the fact that the rate coefficients for this collisional system display a weak temperature dependence in the region from 50 K to 100 K. These low errors thus cannot generally be expected for the extrapolation of rate coefficients corresponding to the excitation of complex organic molecules.

\begin{table}[h!]
\centering
\footnotesize{
{\setstretch{1.5}
\begin{tabular}{c|cc|cc|cc}
& \multicolumn{2}{c}{Extrapolation on $T$}		& \multicolumn{2}{c}{Extrapolation on levels}	& \multicolumn{2}{c}{Combined extrapolation}		\\ 
$T$ (K)	& RMSE	& MEF		& RMSE	& MEF	&  RMSE	& MEF	\\ \hline
600		& 0.093	& 1.18		& 0.129	& 1.24	& 0.208	& 1.47		\\
800		& 0.117	& 1.24		& 0.162	& 1.32	& 0.269	& 1.71		\\
1000		& 0.136	& 1.29		& 0.168	& 1.33	& 0.295	& 1.84		\\
\end{tabular}
}}
\caption{Comparison of different extrapolation schemes for CO-H$_2$ collisional rate coefficients at high temperatures ($T>500$ K).}
\label{table_cross_extrap_CO}
\end{table}

\begin{table}[h!]
\centering
\footnotesize{
{\setstretch{1.5}
\begin{tabular}{c|cc|cc|cc}
& \multicolumn{2}{c}{Extrapolation on $T$}		& \multicolumn{2}{c}{Extrapolation on levels}	& \multicolumn{2}{c}{Combined extrapolation}		\\ 
$T$ (K)	& RMSE	& MEF		& RMSE	& MEF	&  RMSE	& MEF	\\ \hline
60		& 0.062	& 1.11		& 0.113	& 1.22	& 0.127	& 1.25		\\
80		& 0.071	& 1.13		& 0.114	& 1.22	& 0.135	& 1.28		\\
100		& 0.079	& 1.15		& 0.115	& 1.23	& 0.142	& 1.30		\\
\end{tabular}
}}
\caption{Comparison of different extrapolation schemes for \textit{p}-C$_6$H$_5$CN-He collisional rate coefficients at high temperatures ($T>50$ K).}
\label{table_cross_extrap_pBN}
\end{table}

\subsection{Examining the robustness of the model}\label{subsec:dataset_size}

In the preceding sections the extrapolation over rotational levels was performed over the highest 20\% of all levels included in the reference datasets, corresponding to a test set with a size of roughly 60\% of the size of the training set. A natural question is whether the extrapolation results observed for this arbitrary selection of training and test sets are robust when changing the size of the sets.
To investigate this, we focus here exclusively on \textit{p}-C$_6$H$_5$CN-He since the number of rate coefficients in the reference dataset is much larger (5995 rate coefficients at 11 temperatures). 
We examined in detail the behaviour of the extrapolation over unseen rotational levels for varying sizes of the training and test sets. Specifically, the training set included 90\% of all inelastic rate coefficients for transitions between the $N$ first rotational levels, where $N$ varies between 50 and 80 (with 10\% kept for the validation set). We then computed the errors on the predicted rate coefficients for all transitions for which the initial state is within the next 10, 20, 30, or 40 levels.
It should be noted that the number of transitions (and rate coefficients) for a fixed $N$ is given by $N(N-1)/2$, so that the size of the training set does not increase linearly with increasing $N$. 

\begin{figure}
\center
\includegraphics*[width=.95\linewidth]{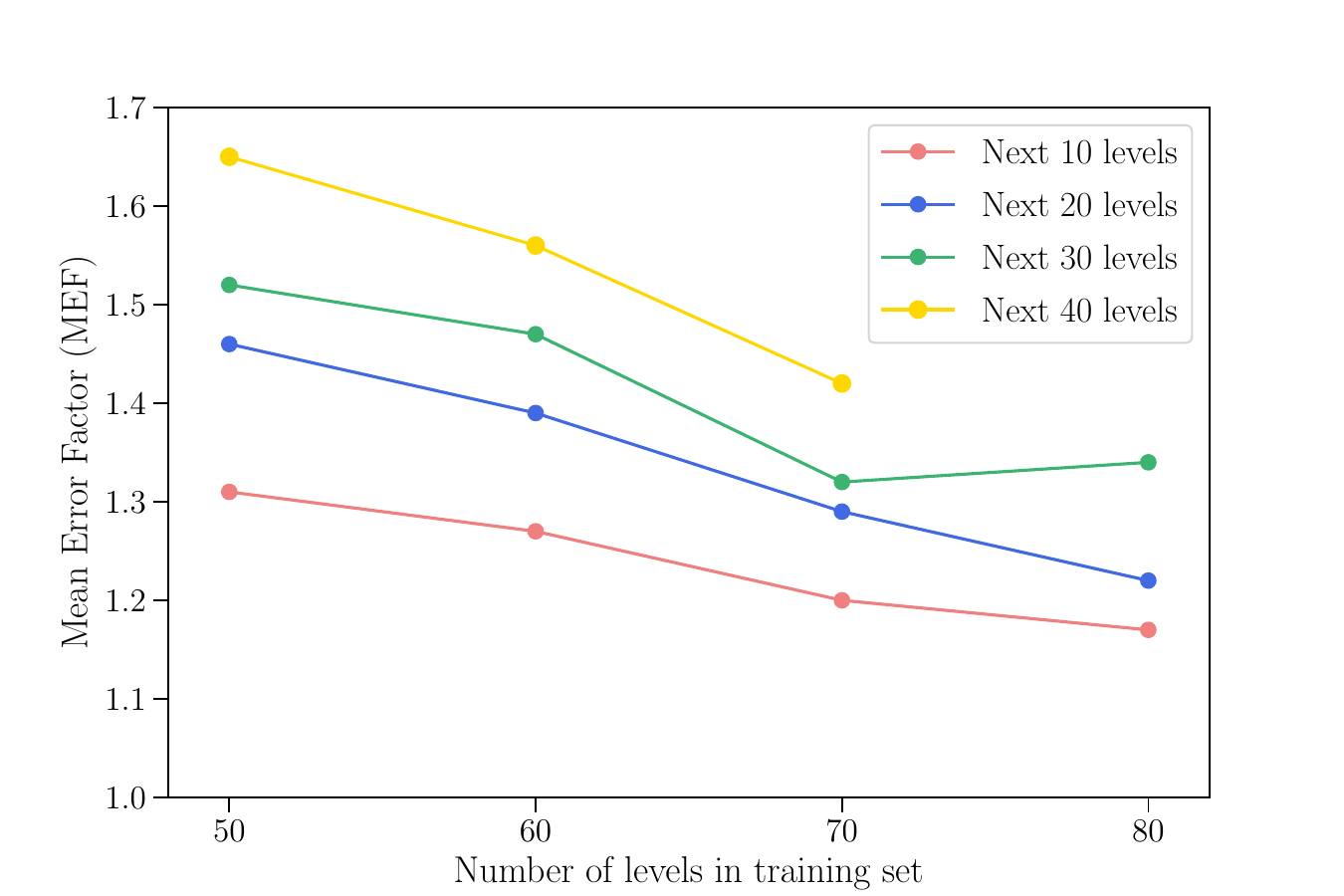}
\caption{Mean error factor (MEF) for various sizes of the training and test sets for \textit{p}-C$_6$H$_5$CN-He. The abscissa corresponds to the number $N$ of rotational levels for which the inelastic rate coefficients are included in the training set. The actual size of the training set is the number of rate coefficients, $N(N-1)/2$ multiplied by the number of temperatures.}
\label{fig:error_size_sets_pBN}
\end{figure}

The results for the MEF (averaged over all transitions and temperatures) are reported in Fig. \ref{fig:error_size_sets_pBN} and Table \ref{table_varying_set_size}. The outcome is aligned with expectations: on the one hand the MEF value decreases with increasing size of the training set since the model is trained on a larger amount of data, but on the other hand the error increases when the extrapolation is made on an increasing number of levels. It is nevertheless worth emphasising that in all cases the model is able to learn the main features of the rate coefficients. As an example, for a model trained on the first 70 levels and extrapolated over the next 30 levels, the test set is slightly larger than the size of the training set. It is therefore possible to double the number of rate coefficients in the reference dataset with a mean error factor around 1.3, which holds promises for astrophysical applications.
The robustness of the model was further tested by computing the standard deviation $\sigma$ based on the MEF on five independent model trainings. The values are reported in Table \ref{table_varying_set_size} for the extrapolation over 20 rotational levels and varying training set size. The standard deviation declines with increasing size of the training set as the model prediction becomes more reliable when the amount of training data expands. 

\begin{table}[h!]
\centering
{\setstretch{1.2}
\begin{tabular}{cccc}
$N$ 	& Size of training set 	& MEF & $\sigma$		\\ \hline
50	& 12127	& 1.46	& 0.11		\\ 
60	& 17523	& 1.39	& 0.10		\\ 
70	& 23908	& 1.29	& 0.07		\\ 
80	& 31284	& 1.22	& 0.02		\\ 
\end{tabular}
}
\caption{MEF for models with varying training set size for \textit{p}-C$_6$H$_5$CN-He, while the extrapolation is always made on transitions involving the next 20 rotational levels. $N$ is the number of levels for which the inelastic rate coefficients are included in the training set, while the second column reports the size of the training set (number of rate coefficients multiplied by the number of temperatures). $\sigma$ is the standard deviation of the MEF based on five models.}
\label{table_varying_set_size}
\end{table}

\subsection{Generalization to other molecular systems}\label{subsec:all_molecules}
The results for CO-H$_2$ and \textit{p}-C$_6$H$_5$CN-He reveal that a neural network model using physically motivated descriptors has the ability to predict state-to-state rate coefficients both for unseen higher-lying rotational levels and in the higher temperature regime despite the different levels of molecular complexity, values of the inelastic rate coefficients, and ranges of temperatures. To further examine the universality of our approach, we tested our model on datasets of de-excitation rate coefficients for several other  molecules relevant to astrophysics and astrochemistry, namely silicon monoxide (SiO), sulfur monoxide (SO), water (H$_2$O), sulfur dioxide (SO$_2$), and vinyl cyanide (CH$_2$CHCN), covering a wide range of molecular properties.
SiO is a molecule present in a wide variety of objects including circumstellar envelopes of AGB stars \cite{Ramstedt2009}.  While SiO is a diatomic closed-shell molecule, its molecular properties differ significantly from those of CO (rotational constant, mass, strength of interaction with H$_2$) so that the trends in collisional excitation rate coefficients contrast with those for CO. A dataset is available \cite{Balanca2018_SiO} for collisions with \textit{p}-H$_2$ containing transitions between 21 energy levels (210 transitions) at 17 temperature in the range of 5-300 K.
SO was chosen as it is an open-shell triplet linear molecule, leading to an increase in complexity. The rotational levels are split by spin-rotation coupling and the levels are characterized by two quantum numbers $j$ and $N$ following the coupling $\bf{j=N+S}$ of the nuclear rotational momentum $\bf{N}$ with the electronic spin angular momentum $\bf{S}$. 
A dataset of inelastic rate coefficients for SO excitation by \textit{p}-H$_2$ is available \cite{Lique2007_SO} and consists of transitions between the lowest 31 levels (465 transitions) with 10 temperatures in the range from 5 K to 50 K.
H$_2$O is one of the most studied molecules in astrochemistry, with a high abundance in molecular clouds and star-forming regions. Due to its importance in astrophysical environments, it was also originally used to test the performance of a neural network model on predictions of inelastic rate coefficients by Neufeld \cite{Neufeld2010_ANN}, while Faure and Josselin applied an information theory approach to extrapolate rovibrational rate coefficients \cite{Faure2007_H2O}. As an asymmetric top, its rotational levels are determined by the three quantum numbers $j$, $k_a$, and $k_c$. It also exists in the form of two spin isomers (\textit{para} and \textit{ortho}) that do not interconvert in inelastic collisions. The dataset \cite{Daniel2011_H2O} used includes transitions among the first 45 levels of both \textit{ortho}-H$_2$O (hereafter \textit{o}H$_2$O) and \textit{para}-H$_2$O (hereafter \textit{p}H$_2$O) following collisions with \textit{p}H$_2$ (990 transitions for each case) over 13 temperatures values ranging from 5~K to 1500~K.
SO$_2$ has a much denser energy level structure than H$_2$O, leading to a different behaviour for its inelastic rate coefficients. The dataset utilized for this system \cite{Cernicharo_SO2} encompasses de-excitation transitions between first 31 rotational levels (465 transitions) across 6 different temperatures ranging from 5 to 30 K. 
Finally, vinyl cyanide (CH$_2$CHCN) 
provides an additional example of COMs displaying a high density of rotational levels. Some rotational transitions of CH$_2$CHCN exhibit maser features \cite{Gardner1975_vinylcyanide} that require non-LTE models to elucidate, and rate coefficients for excitation by He atoms are available \cite{Sogomonyan2025a} for transitions among the lowest 110 levels (5995 transitions) in the range 5-50 K. 

\begin{figure}
\center
\includegraphics*[width=1\linewidth]{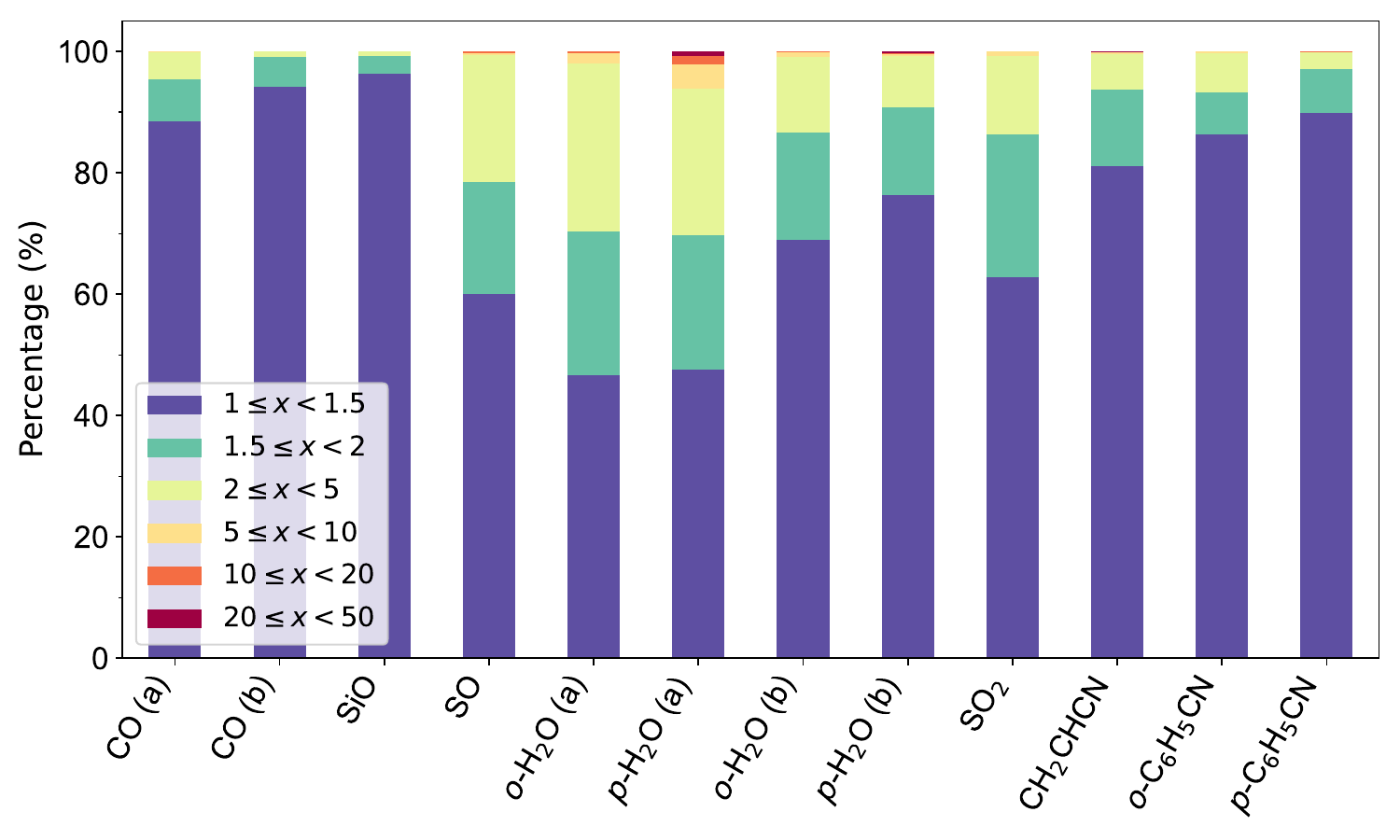}
\caption{Fraction (in \%) of all predicted rate coefficients with an error factor $x$ in a certain range for all molecules reported in Table \ref{table_all_molecules}. ``CO (a)'' corresponds to the model trained and tested on the full temperature range ($T=2-3000$ K), while for ``CO (b)'' the temperature range is limited to $T\geq 20$ K. ``H$_2$O (a)'' corresponds to the model trained and tested the full temperature range ($T=5-1500$ K), while for ``H$_2$O (b)'' the temperature range is limited to $T>200$ K. See also Table \ref{table_all_molecules}.}
\label{fig:error_fraction_all_mols}
\end{figure}

NN models were trained for each molecule for an extrapolation over de-excitation rate coefficients involving the highest 20\% of levels. 
The results of the model prediction for these molecules are summarized in Table \ref{table_all_molecules} along with the features of the datasets, while Fig. \ref{fig:error_fraction_all_mols} shows the percentage of predicted rate coefficients within fixed ranges of error factors for all molecules. Additional comparisons between predicted and actual rate coefficients are presented in Figs. S4-S6 of the Supplementary Material. 

SiO performs exceedingly well, with a MEF of only 1.13 and a WMEF of 1.07. The better performance compared to CO can likely be attributed to the smaller range of temperature as well as to the smaller range of values taken by the rate coefficients. We tested this hypothesis by re-training a model on the CO rate coefficients without the lowest temperatures ($T < 20$ K), a region where trends in rate coefficients might be harder to capture due to the influence of quantum effects. This leads to an improvement of the error measures and a performance comparable to that seen for SiO. 
The errors are larger for SO, as expected due to the presence of the fine structure (and the associated additional quantum number) and the limited size of the training set. The much smaller value of the WMEF compared to the MEF (1.28 vs. 1.68) shows that the dominant rate coefficients are well reproduced, although 21\% of all rate coefficient show a deviation larger than a factor 2 and 0.6\% show a deviation larger than a factor 5 (see Fig. \ref{fig:error_fraction_all_mols}). 
H$_2$O displays the worst results, despite a large training set, with MEFs of 1.88 for \textit{o}H$_2$O and 2.32 for \textit{p}H$_2$O and a fraction of transitions affected by an error up to a factor 50 for \textit{p}H$_2$O. We attribute these large errors to two factors: the large temperature range covered in the dataset (5-1500 K), and the absence of simple propensity rules for the excitation process in H$_2$O-H$_2$ collisions \cite{Faure2007_H2O}, especially at low temperature. As in the case of CO discussed above, the first hypothesis can be tested by restricting the range of temperatures in the training set. Removing temperatures below 200 K leads to a dramatically improved MEF of 1.53 for \textit{o}H$_2$O and 1.51 for \textit{p}H$_2$O, and eliminates the largest outliers (see Fig. S7 of the Supplementary Material). The second hypothesis is more difficult to test, but a comparison with the case of SO$_2$ is informative as the collisional rate coefficients for SO$_2$-\textit{p}H$_2$ display a clear propensity rule for transitions which conserve the value of $k_a$ ($\Delta k_a =0$). The error measures are indeed lower than for H$_2$O-\textit{p}H$_2$, despite the much smaller size of the training set.
The possibility of predicting H$_2$O-\textit{p}H$_2$ rate coefficients using neural network architectures was previously investigated by Neufeld in a seminal work \cite{Neufeld2010_ANN} at a fixed temperature of 800 K. The best model gave a RMSE $\log_{10}$ of 0.41 for \textit{o}H$_2$O-\textit{p}H$_2$, while we find a much lower value of 0.21. This coincides with a fraction of transitions with an error larger than a factor 10 that is also much lower.

Finally, the results for CH$_2$CHCN show that the low errors observed in the case of C$_6$H$_5$CN are not accidental, with an overall MEF of 1.34 and a WMEF of 1.2, and a fraction of predicted rate coefficients with an error larger than a factor 5 below 0.5\%. As discussed in the the comparison of H$_2$O and SO$_2$, a possible explanation lies in the strong $\Delta k_a=0$ propensity rules for the excitation of CH$_2$CHCN and C$_6$H$_5$CN \cite{Sogomonyan2025a,BenKhalifa2024a}, related to the near-prolate character of these molecules \cite{Sogomonyan2025b}, that are more easily captured by the NN model. 

The comparison of models extrapolating rate coefficients to transitions involving unseen rotational levels for a wide variety of molecules thus reveal several trends. First, the model architecture is capable of extrapolating with an overall error lower than a factor 2, and for most molecules lower than a factor 1.5. Second, when the dataset covers a large range of temperatures, the model performance can usually be improved by reducing this range, as observed for the cases of CO and H$_2$O. Third, while the best results were found for diatomic molecules, an increase in molecular complexity (number of atoms, density of rotational states) does not necessarily correlate with a decrease in the model's performance. The predictive capability of the model is driven by an interplay between the size of the training set, the molecular complexity (and the associated quantum numbers used as descriptors), and the propensities in the collisional rate coefficients. Finally, WMEF values are consistently lower than MEF values, indicating that the error is smaller for the larger rate coefficients, which is a favorable outcome for applications to radiative transfer models in astrochemistry.

\begin{table*}[htbp]
\centering
{\setstretch{1.5}
\footnotesize{
\begin{tabular}{c|cccccccccccc}
Molecule			& CO (a)		& CO (b)		& SiO	& SO	& $o$-H$_2$O (a)	& $p$-H$_2$O (a)	& $o$-H$_2$O (b)	& $p$-H$_2$O (b)	& SO$_2$	& $o$-C$_6$H$_5$CN	& $p$-C$_6$H$_5$CN	& CH$_2$CHCN	\\ \hline
Levels in dataset	& 41			& 41			& 21		& 31		& 45		& 45		& 45		& 45		& 31		& 104		& 110	& 110		\\
Range of $T$ (K)	& $2-3000$	& $20-3000$	& $5-300$	& $5-50$	& $5-1500$		& $5-1500$	& $200-1500$	& $200-1500$	& $5-30$	& $5-100$	& $5-100$	& $5-50$		\\
Temperatures		& 25			& 22			& 17		& 10		& 13		& 13		& 8		& 8		& 6		& 11		& 11		& 10				\\
Size training set	& 11880		& 10454		& 2081	& 2700	& 7371	& 7371	& 4536	& 4536	& 1620	& 32878	& 37036	& 34452				\\
Size validation set	& 1320		& 1162		& 231	& 300	& 819	& 819	& 504	& 504	& 180	& 3653	& 4115	& 3828				\\
Size test set		& 7300		& 6424		& 1258	& 1650	& 4680	& 4680	& 2880	& 2880	& 990	& 22385	& 24794	& 21670				\\ \hline
RMSE			& 0.131		& 0.090		& 0.073	& 0.249	& 0.292	& 0.357	& 0.210	& 0.199	& 0.217	& 0.151	& 0.119	& 0.158		\\
Median			& 0.019		& -0.003		& 0.001	& 0.044	& -0.128	& -0.117	& -0.06	& -0.06	& -0.056	& -0.009	& 0.01	& 0.002		\\
MEF				& 1.25		& 1.17		& 1.13	& 1.68	& 1.88	& 2.32	& 1.53	& 1.51	& 1.56	& 1.30	& 1.23	& 1.34		\\
WMEF			& 1.095		& 1.13		& 1.07	& 1.28	& 1.69	& 2.51	& 1.37	& 1.32	& 1.45	& 1.27	& 1.18	& 1.20		\\
$1\leq x \leq 1.5$	& 88.55		& 94.13		& 96.35	& 60		& 46.62	& 47.64	& 68.96	& 76.39	& 62.83	& 86.32	& 89.9	& 81.17		\\
$1.5 < x \leq 2$		& 6.92		& 4.92		& 2.86	& 18.55	& 23.78	& 22.05	& 17.67	& 14.37	& 23.54	& 6.97	& 7.22	& 12.59		\\
$2<x\leq 5$		& 4.37		& 0.95		& 0.79	& 20.85	& 27.63	& 24.29	& 12.47	& 8.47	& 12.93	& 6.44	& 2.76	& 5.82		\\
$5<x\leq 10$		& 0.16		& 0			& 0		& 0.36	& 1.67	& 3.89	& 0.76	& 0.42	& 0.71	& 0.27	& 0.11	& 0.36		\\
$10<x\leq 20$		& 0			& 0			& 0		& 0.24	& 0.3	0	& 1.47	& 0.14	& 0.07	& 0		& 0		& 0.01	& 0.05		\\
$20<x\leq 50$		& 0			& 0			& 0		& 0		& 0		& 0.66	& 0		& 0.28	& 0		& 0		& 0		& 0.01		\\
\end{tabular}
}}
\caption{Summary of the datasets used for each molecule, overall error measures (averaged over the temperature range and over the extrapolated rate coefficients), and fraction (in \%) of all predicted rate coefficients with an error factor $x$ in a given range.}
\label{table_all_molecules}
\end{table*}

\section{Conclusions}\label{sec:conclusions}

We have explored the use of a multilayer perceptron model to predict state-to-state collisional rate coefficients for a broad range of molecular systems relevant to astrophysics and astrochemistry. By combining physically motivated descriptors, including quantum numbers, level energies, energy gaps, and kinetic temperature, the model successfully learned the main trends governing inelastic molecular collisions and generalized to regions of parameter space not included in the training data.
The results show that extrapolation toward higher rotational levels is feasible with modest errors. For most molecules investigated, the predicted rate coefficients differed from reference calculations by less than a factor of 1.5 on average, while even the most challenging cases generally remained within a factor of two. The model was also able to extrapolate toward higher temperatures, often with greater accuracy than for level extrapolation, reflecting the smooth temperature dependence of many collisional rate coefficients. Simultaneous extrapolation in both temperature and rotational excitation represented the most demanding scenario but still produced errors comparable to, or smaller than, those commonly associated with approximate scattering approaches. 

A comparison across molecular systems revealed several important trends. The predictive performance depends not only on molecular complexity but also on the size of the training dataset and on the presence of clear propensity rules governing collisional transitions. Molecules such as SiO, benzonitrile, and vinyl cyanide exhibited particularly good performance, whereas larger errors were obtained for H$_2$O, whose collisional dynamics are characterized by more complex propensity patterns. We further found that excessively broad temperature ranges can degrade model performance at low temperatures and that restricting training to more homogeneous temperature domains can substantially improve prediction accuracy. 
The influence of the size of the training set was examined in detail for the excitation of CO and C$_6$H$_5$CN, confirming the expectations that the accuracy increases with increasing size of the training set but decreases when the extrapolation is made on an increasing number of levels. These examples suggest that a doubling of the reference dataset of rate coefficients is feasible with errors of the order of a factor 1.5, although the precision depends on the molecule considered.

These findings demonstrate that machine learning can serve as a valuable complement to traditional quantum scattering calculations. In particular, neural-network models offer a practical and computationally inexpensive route of expanding existing collisional datasets to higher rotational levels and temperatures beyond the limits of current quantum scattering calculations. This capability is especially valuable for radiative transfer modeling of warm astrophysical environments including star-forming regions, circumstellar envelopes, and protoplanetary disks.
From an astrophysical perspective, these results are particularly encouraging since the calculation of extensive and accurate collisional datasets remains one of the major bottlenecks in the interpretation of molecular spectra under non-LTE conditions, especially for complex organic molecules and highly excited rotational states. 

Future work should investigate more advanced architectures, incorporate uncertainty quantification, and explore transfer-learning strategies capable of taking advantage of information from multiple molecular systems simultaneously. Such developments may ultimately enable the generation of reliable collisional data for complex molecules whose direct quantum-mechanical treatment remains computationally inaccessible, such as the excitation of COMs by H$_2$. 

\section*{Supplementary Material}
The Supplementary Material file contains additional comparisons of model predictions of rate coefficients for the cases presented in Sec. \ref{subsec:extrap_levels} and \ref{subsec:all_molecules}.

\begin{acknowledgments}
JL acknowledges support from KU Leuven through Project no. C14/22/082. 
\end{acknowledgments}

\section*{Appendix: level list for the \textit{para}-C$_6$H$_5$CN dataset.}
\begin{table*}[h!]
\centering
\begin{small}
{\setstretch{1.0}
\begin{tabular}{ccc|ccc|ccc}
Level& $j_{k_a, k_c}$	& Energy (cm$^{-1}$)	& Level& $j_{k_a, k_c}$	& Energy (cm$^{-1}$) & Level& $j_{k_a, k_c}$	& Energy (cm$^{-1}$) 	\\ \hline
1	&	$1_{1,1}$		&	0.2291	& 41	&	$7_{5,2}$		&	6.1481	& 81	&	$10_{7,3}$		&	12.0611	\\
2	&	$1_{1,0}$		&	0.2402	& 42	&	$10_{3,8}$	&	6.3775	& 82	&	$15_{3,13}$	&	12.3887	\\
3	&	$2_{1,2}$		&	0.4023	& 43	&	$10_{3,7}$	&	6.4142	& 83	&	$15_{3,12}$	&	12.6646	\\
4	&	$2_{1,1}$		&	0.4355	& 44	&	$11_{1,10}$	&	6.4452	& 84	&	$16_{1,15}$	&	12.7867	\\
5	&	$3_{1,3}$		&	0.6616	& 45	&	$12_{1,12}$	&	6.7863	& 85	&	$17_{1,17}$	&	13.0632	\\
6	&	$3_{1,2}$		&	0.7281	& 46	&	$8_{5,4}$		&	6.8881	& 86	&	$11_{7,5}$		&	13.0784	\\
7	&	$4_{1,4}$		&	1.0066	& 47	&	$8_{5,3}$		&	6.8881	& 87	&	$11_{7,4}$		&	13.0784	\\
8	&	$4_{1,3}$		&	1.1173	& 48	&	$11_{3,9}$	&	7.3976	& 88	&	$14_{5,10}$	&	13.294	\\
9	&	$5_{1,5}$		&	1.4368	& 49	&	$11_{3,8}$	&	7.4591	& 89	&	$14_{5,9}$		&	13.2958	\\
10	&	$5_{1,4}$		&	1.6025	& 50	&	$12_{1,11}$	&	7.5543	& 90	&	$16_{3,14}$	&	13.8597	\\
11	&	$3_{3,1}$		&	1.8363	& 51	&	$9_{5,5}$		&	7.7213	& 91	&	$12_{7,6}$		&	14.1889	\\
12	&	$3_{3,0}$		&	1.8363	& 52	&	$9_{5,4}$		&	7.7213	& 92	&	$12_{7,5}$		&	14.1889	\\
13	&	$6_{1,6}$		&	1.9516	& 53	&	$13_{1,13}$	&	7.8792	& 93	&	$16_{3,13}$	&	14.2204	\\
14	&	$6_{1,5}$		&	2.1826	& 54	&	$12_{3,10}$	&	8.5096	& 94	&	$17_{1,16}$	&	14.2922	\\
15	&	$4_{3,2}$		&	2.206	& 55	&	$12_{3,9}$	&	8.6062	& 95	&	$18_{1,18}$	&	14.562	\\
16	&	$4_{3,1}$		&	2.2061	& 56	&	$10_{5,6}$	&	8.6479	& 96	&	$15_{5,11}$	&	14.6919	\\
17	&	$7_{1,7}$		&	2.5504	& 57	&	$10_{5,5}$	&	8.648	& 97	&	$15_{5,10}$	&	14.6953	\\
18	&	$5_{3,3}$		&	2.6685	& 58	&	$13_{1,12}$	&	8.7438	& 98	&	$13_{7,7}$		&	15.3925	\\
19	&	$5_{3,2}$		&	2.6691	& 59	&	$14_{1,14}$	&	9.0534	& 99	&	$13_{7,6}$		&	15.3925	\\
20	&	$7_{1,6}$		&	2.8561	& 60	&	$7_{7,0}$		&	9.5664	& 100	& $17_{3,15}$	&	15.4182	\\
21	&	$6_{3,4}$		&	3.224	& 61	&	$7_{7,1}$		&	9.5664	& 101	&	$9_{9,0}$		&	15.695	\\
22	&	$6_{3,3}$		&	3.226	& 62	&	$11_{5,7}$	&	9.6682	& 102	&	$9_{9,1}$		&	15.695	\\
23	&	$8_{1,8}$		&	3.2327	& 63	&	$11_{5,6}$	&	9.6683	& 103	&	$17_{3,14}$	&	15.8747	\\
24	&	$8_{1,7}$		&	3.6214	& 64	&	$13_{3,11}$	&	9.7127	& 104	&	$18_{1,17}$	&	15.877	\\
25	&	$7_{3,5}$		&	3.8728	& 65	&	$13_{3,10}$	&	9.8564	& 105	&	$16_{5,12	}$	&	16.1848	\\
26	&	$7_{3,4}$		&	3.8776	& 66	&	$14_{1,13}$	&	10.0127	& 106	&	$16_{5,11}$	&	16.1911	\\
27	&	$9_{1,9}$		&	3.9978	& 67	&	$8_{7,2}$		&	10.3052	& 107	&	$10_{9,1}$		&	16.6184	\\
28	&	$9_{1,8}$		&	4.4764	& 68	&	$8_{7,1}$		&	10.3052	& 108	&	$10_{9,2}$		&	16.6184	\\
29	&	$8_{3,6}$		&	4.6147	& 69	&	$15_{1,15}$	&	10.3089	& 109	&	$14_{7,8}$		&	16.6897	\\
30	&	$8_{3,5}$		&	4.6252	& 70	&	$12_{5,8}$	&	10.7825	& 110	&	$14_{7,7}$		&	16.6897	\\
31	&	$10_{1,10}$	&	4.8453	& 71	&	$12_{5,7}$	&	10.7828	\\
32	&	$5_{5,1}$		&	4.9468	& 72	&	$14_{3,12}$	&	11.006	\\
33	&	$5_{5,0}$		&	4.9468	& 73	&	$9_{7,3}$		&	11.1368	\\
34	&	$10_{1,9}$	&	5.4185	& 74	&	$9_{7,2}$		&	11.1368	\\
35	&	$9_{3,7}$		&	5.4497	& 75	&	$14_{3,11}$	&	11.2094	\\
36	&	$9_{3,6}$		&	5.4701	& 76	&	$15_{1,14}$	&	11.3603	\\
37	&	$6_{5,2}$		&	5.5011	& 77	&	$16_{1,16}$	&	11.6455	\\
38	&	$6_{5,1}$		&	5.5011	& 78	&	$13_{5,9}$	&	11.991	\\
39	&	$11_{1,11}$	&	5.7749	& 79	&	$13_{5,8}$	&	11.9918	\\
40	&	$7_{5,3}$		&	6.1481	& 80	&	$10_{7,4}$	&	12.0611	\\
\end{tabular}
}
\end{small}
\caption{Level list for the \textit{para}-C$_6$H$_5$CN dataset.}
\label{table_levels_pBN}
\end{table*}

\bibliography{bib_ML_paper}
\end{document}